\documentclass[pre,reprint]{revtex4-1}
\usepackage{graphicx}
\usepackage[centertags]{amsmath}
\usepackage{amsfonts}
\usepackage{amssymb}
\usepackage{amsthm}
\usepackage{newlfont}
\usepackage{stmaryrd}
\usepackage{mathrsfs}
\usepackage{euscript}
\usepackage{braket}
\usepackage{bbm}
\usepackage{natbib}
\usepackage{lipsum}

\usepackage{color}

\draft % marks overfull lines with a black rule on the right

\begin{document}

%\preprint{}

\title{From Langevin to generalized Langevin equations\\ for the nonequilibrium Rouse model} %Title of paper



\author{Christian Maes}
%\email[]{christian.maes@fys.kuleuven.be}
%\thanks{}
\affiliation{Instituut voor Theoretische Fysica, KU Leuven,
Belgium}


\author{Simi R. Thomas}
\email[]{simi.thomas@fys.kuleuven.be}
%\thanks{}
\affiliation{Instituut voor Theoretische Fysica, KU Leuven,
Belgium}

\date{\today}

\begin{abstract}
We investigate the nature of the effective dynamics and statistical forces obtained after integrating out nonequilibrium degrees of freedom. To be explicit, we consider the Rouse model for the conformational dynamics of an ideal polymer chain subject to steady driving.    We compute the effective dynamics for one of the many monomers by integrating out the rest of the chain.  The result is a generalized Langevin dynamics for which
we give the memory and noise kernels and the effective force, and we discuss the inherited nonequilibrium aspects.
\end{abstract}


\maketitle %\maketitle must follow title, authors, abstract and \pacs

\section{Introduction}

To relate different levels of physical description belongs to the core business of statistical mechanics. That is accompanied by the emergence of notions as heat and entropy, that are important concepts for bridging the 
gap between microscopic laws and macroscopic behavior.  
And also ``new'' forces can appear, effectively from integrating out certain degrees of freedom. These are commonly called entropic or statistical forces and they have been discussed since relaxation to equilibrium was first 
considered. Indeed, for equilibrium dynamics, statistical forces have their origin in the nature of macroscopic systems to evolve towards higher entropy rather than being caused by some specific mechanical force.  Though much 
systematic progress has been made in the study of these entropic forces, there has been less venture into the effects of nonequilibrium driving on statistical forces. Only over the last decade systematic efforts have been made
to calculate these nonequilibrium statistical effects. That includes Soret-Casimir forces \cite{naj,krug} but also general considerations on the law of action and reaction \cite{sas}.

The first motivation of the present work is 
to study an explicit example of statistical forcing 
emerging from integrating out a nonequilibrium environment. Yet, the case we treat comes with an extra motivation as it opens some questions in the nonequilibrium physics of polymers. In contrast to many ongoing studies of 
nonequilibrium polymer rheology, of transport through polymers 
or of mechanical folding and stretching of polymers,  the present paper considers also steady nonequilibria, i.e., where the driving is constant in time and the condition of detailed balance is broken.\\
  Our working model is the widely studied Rouse model  \cite{rousetheory}, an ideal chain, where monomers are connected through Gaussian springs, and excluded volume effects and hydrodynamic interactions are neglected. 
This model holds a special place, as it is the simplest model which can be exactly solved to describe phenomena like anomalous diffusion
of polymers in a bath. Moreover, in the natural context of polymer melts, which are a collection of polymers in a solution, the diffusion of a tagged polymer can be described by Rouse dynamics moving along a one-dimensional 
tube embedded
in a network or mesh of polymers \cite{doitheory,de1scaling}. Our general question can then be asked here, to investigate the effective dynamics of a tagged monomer when the chain is subjected to nonequilibrium driving. We
have in mind that the extremal monomers are subject to non-conservative forces e.g. via a small charged particle or optically driven bead attached to them, and we wish to follow a tagged monomer near the middle of the chain.\\

Because of the harmonic interaction, the nonequilibrium Rouse model is one of the simplest, still physically interesting examples to understand the effective dynamics in a driven medium.
The equilibrium version of integrating out the Rouse model was already carried out by D.~Panja \cite{panja2010}.  The dynamics of the tagged monomer is shown to be non-Markovian with memory kernel having a power law decay $\mu(t) \propto t^{-1/2}$
for short times and exponential decay asymptotically in time,
\begin{equation}
 \mu(t) \propto \frac 1{\sqrt{t}}\, e^{-t/\tau}
\end{equation}
The kernel $\mu(t)$ is shown to be the mean relaxation response of the polymers to local strain and its behavior gives good information on the nature of the diffusion, which is anomalous for intermediate times, $\Delta x^2 \propto D \sqrt{t}$.\\
In the present paper we start with the phantom Rouse dynamics
in the inertial regime and we introduce a nonequilibrium driving. The result of integrating out the (other) polymer degrees of freedom is again a generalized Langevin equation (GLE) for the tagged monomer. We show that
in the overdamped limit, the equilibrium results match those of Panja \cite{panja2010}. We discuss the nonequilibrium corrections to the force and memory terms for driving of specific nature to obtain some general information about statistical forces in nonequilibrium.\\
The more systematic and general approach to integrating out degrees of freedom is commonly referred to as the Mori-Zwanzig approach \cite{zwanzigmemory,moritransport} or the approach via adiabatic elimination \cite{gardiner,mackay,berrychaotic,jarzy}.
Generalized Langevin equations have also been derived in nonstationary environments \cite{kaw} and similar in spirit to the present paper is also the generalization  where a coarse-graining is added upon a coarse-grained
description \cite{esp}, or  where one Brownian particle is described in a nonequilibrium bath \cite{shea}.  In the case of nonequilibrium thermostated dynamics a generalized Langevin equation has also been derived \cite{eva}. 
 We do not follow these general schemes here also because we work on the more explicit Langevin (not Fokker-Planck) side of the question, and we take no special limits for macroscopic systems or for the speed of motion of non-conserved versus 
conserved quantities. Moreover, these general approaches are less explored for starting with open driven polymer dynamics as we do here. An interesting study of the average square displacement of a tagged monomer in Rouse polymer chain subjected to  random, layered convection flows both time-independent and time-dependent has been made in \cite{oshanin1994rouse, jespersen2001polymer}. \\

In the next section we introduce the model and the various types of nonequilibrium driving.
Sections \ref{resu}, \ref{she-dimer} and \ref{trap} summarize the method and the results with a discussion of the effective dynamical behavior.  Finally some more essential elements of the computations are collected in Appendix  \ref{pros}.
 
\section{Nonequilibrium Rouse dynamics}

We consider the positions $\vec{R}_i, i=1,\ldots, N,$ of $N$ point particles (also called monomers) in a three-dimensional domain open to thermal exchanges.
The particles are harmonically coupled and some are subject to further forces, some of which are non-conservative. The potential energy is quadratic
\begin{equation}\label{us}
U(\vec{R}) = \frac{\kappa}{2}(\vec{R}_1-\vec{R}_2)^2 + \frac{\kappa}{2}(\vec{R}_2-\vec{R}_3)^2  +\ldots + \frac{\kappa}{2}(\vec{R}_{N-1}-\vec{R}_N)^2
\end{equation}
and the force on the $i$th particle is the sum of systematic forces $\vec{K}_i$ and Langevin forces $\vec{L}_i$:
\begin{eqnarray}
\vec{K}_i = -\vec{\nabla}_i U + \vec{F}_i,\quad \vec{L}_i = -m\gamma \dot{\vec{R}}_i + \vec{\xi}_i
\end{eqnarray} 
There is an independent standard white noise $\vec{\xi_i}$ modeling the action of the thermal environment at temperature $T$ and friction $\gamma$, with $\langle \vec{\xi}_{i,\alpha}(t) \vec{\xi}_{j,\beta}(t')\rangle = 2m\gamma k_BT\delta_{i,j}\delta_{\alpha,\beta} \delta(t-t')$, where $\alpha$ and $\beta$
refer to the various spatial directions.
The first term in $\vec{K}_i$ is the conservative part of the force. The force $\vec{F}_i$ need not be conservative or constraining and will be specified below; that is what we refer to as the driving.   
We then have the equation of motion for the time-dependent coordinates $\vec{R}_i(t)$
\begin{equation}\label{eq2}
 m\frac{d^2 \vec{R}_i}{dt^2} = \vec{K_i} + \vec{L}_i
 \end{equation}
 with given initial conditions $R_i(0), \dot{\vec{R}}_i(0)$ at time $t=0$. In many cases of standard polymer physics the inertial term proportional to the mass in \eqref{eq2} can be fairly ignored.  That can be done in all following equations and results but there is however no harm in keeping it; in fact our concern is not in the first place towards a detailed study in polymer physics.  In fact, we take the Rouse polymer model for the simple purpose of illustrating effects of statistical forces in a nonequilibrium environment.  To have a workable model we can exploit the linearity of the Rouse model and the extra forcing $\vec{F}_i$ will also be assumed linear.   We will however not proceed with a diagonalization, and we will not write the solution in terms of modes.  After integrating out all particles but the first one, we  obtain explicit 
information about the final equation of the form,\\

\begin{widetext}
\begin{equation}\label{gle}
 m\frac{d^2 \vec{R}_1}{dt^2} =  -m\int_0^t dt' \mu^{(N)}(t-t') \dot{\vec{R}}_1(t') - m\gamma \dot{\vec{R}}_1(t) + \vec{\eta}^{(N)}(t) + \vec{\xi}_1(t)+ \vec{G}^{(N)}(t)
\end{equation}
\end{widetext}

Indeed, not surprisingly and as an explicit example of a type of Zwanzig's program \cite{zwanzignonequilibrium}, we will find the validity of a GLE of  the form \eqref{gle}.  Our model will enable rather explicit memory and friction kernels.
 We will discuss the memory kernel $\mu(t)$ (more generally a matrix), the noise $\vec{\eta}(t)$ and the statistical forcing $\vec{G}$ (that all depend on $N$) in the cases that we introduce next. %\textbf{Is it important to say/discuss the difference with going for the middle polymer??}
 Obviously, the case of the effective dynamics on another coordinate, e.g. the middle one around $i=N/2$, can be reduced to that case.
 The effective force $\vec{G}$ can be of convolution type, as in Eq. \eqref{force1x} below, and also contain the memory of the past trajectory of the tagged monomer. 

\vspace{0.25pt} 

\subsection{Uniform constant driving}\label{intor}
The simplest case is to assume  that the outer end of the polymer is being driven under a constant external force $f$.  That is a mathematical idealization of a polymer say with a charged end, forced under an electric field.  
As there is no confining force for the polymer, that means the whole system will move in the direction of the field and we discuss that diffusive regime.   That is, we take free boundary conditions and $\vec{F}_i = \delta_{N,i}\,f\, \hat{e}_x$, for some constant field $f$ in the $x-$direction.  The simplest example corresponds to two linearly coupled degrees of freedom moving in one dimension, with dynamics
\begin{eqnarray}\label{cons1}
m\frac{d^2 R_1}{dt^2}  &=& -\kappa [R_1 -R_2] - m\gamma   \frac{d}{dt} R_1 + \xi_1(t) \nonumber\\
m\frac{d^2 R_2}{dt^2} & =& -\kappa [R_2 -R_1] - m\gamma \frac{d}{dt} R_2 +
\xi_2(t)  + f
\end{eqnarray} 
for $R_1,R_2 \in \mathbb{R}$.
The constant $f$ induces a drift.  We give here that dimer-case explicitly also because we have found that for all finite $N$ (size of original polymer) the basic qualitative features of generalized memory and friction are unchanged 
from $N=2$, where things are of course much simpler.

%  \begin{figure}[ht!]
%  \begin{center}
%  \includegraphics[totalheight=7cm,width=9cm]{long_time.epsi}%,width=5in
%  \caption{\label{fig3.1}The long time limit of memory in a semi-log plot. The dotted curves are the analytic values and the full curves are linear fits. 
% The slope of the various curves is seen to be inversely proportional to $N^2$ ($N$ is the system size).}
%  \end{center}
%  \end{figure}

\subsection{Non-uniform driving}\label{she}
Here we imagine the motion of a polymer in a 2-dimensional slab of vertical size $L$ in which the outer end is subject to a forcing in the horizontal direction that is linear in the vertical distance.  We can imagine that as the result of a shearing at the outer edge of the polymer, but we do not imagine a surrounding fluid as we wish to stick to the Rouse model (ignoring hydrodynamic interactions as e.g. in the Zimm model).  
In terms of a polymer melt we can realize that by attaching a bead or nanoparticle to the end of the polymer chain, which is then driven in one direction but non-uniformly with respect to an orthogonal direction.  That provides a well known case of a non-conservative force.\\
For explicitness we write out this case again first for a polymer of size $N=2$.  One monomer is being acted on by the non-uniform force which depends on its $y$-coordinate. The equation of motion written in Cartesian coordinates is then
\begin{eqnarray}\label{shear1}
 \nonumber
\frac{md^2R_{2x}}{dt^2} &=& k[R_{1x} -R_{2x}] - m\gamma \frac{dR_{2x}}{dt} + \xi_{2x}(t) + f\,R_{2y}\\
\nonumber
\frac{md^2R_{2y}}{dt^2} &=& k[R_{1y} -R_{2y}] - m\gamma \frac{dR_{2y}}{dt} +\xi_{2y}(t) \\
\nonumber
\frac{md^2R_{1x}}{dt^2} &=& -k[R_{1x} -R_{2x}] - m\gamma \frac{dR_{1x}}{dt} +\xi_{1x}(t) \\
\frac{md^2R_{1y}}{dt^2} &=& -k[R_{1y} -R_{2y}] - m\gamma \frac{dR_{1y}}{dt} +\xi_{1y}(t)
\end{eqnarray}
where $f$ is the nonequilibrium amplitude.\\  

Since the external force now does depend on the position, it is useful here to have a comparison or equilibrium reference, where an external potential $U_{\text{ext}}$ is added to the potential energy $U$ so to trap the outer monomer.  In other words, again for simplicity of presentation, for the case of a dimer, 
$F(t) = -f(R_2 -Q)$  
which derives from a confining potential around position $Q$ which holds the outer edge of the polymer.
\begin{eqnarray}\label{trap1}
\nonumber
  m\frac{d^2 R_2}{dt^2} &=& -\kappa(R_2 - R_1) -m\gamma \frac{d R_2}{dt} + \xi_2(t) -f(R_2-Q)\\
m\frac{d^2 R_1}{dt^2} &=& -\kappa(R_1 -R_2) -m\gamma \frac{d R_1}{dt} +\xi_1(t)
\end{eqnarray} which would replace the dynamics of the $x$-components of the equations \eqref{shear1}.\\

\section{General method: induction and recurrence relations}
In this section we show the methods and intermediate steps involved in reaching our results that will be summarized in the next three sections.  The general method is always to work via iteration and to prove results by induction.
More precisely, the tagged particle equation of motion is directly coupled to a second particle which then is coupled to the other $N-2$ particles.  If we now assume  that first, 
after integrating out these  $N-2$ particles, the  effective dynamics on the second particle is of the form \eqref{gle}, then we obtain two equations: one is the GLE \eqref{gle} 
(with $N$ there replaced by $N-1$) and the other is the original equation of motion of the tagged particle coupled to the second particle.  Assuming the structure \eqref{gle} for $N-1$
with specific properties of the memory kernel, noise and force constitutes the induction hypothesis.  The remaining task is then to integrate out that last (second) particle and to prove 
that the induction hypothesis is indeed reproduced at size $N$.  The crucial step to discover what is the correct induction hypothesis is the case $N=2$.  That is also why the essential first step is to be explicit about the case $N=2$.  We next give more details.\\

After integrating out $N-2$ particles we arrive at the following GLE for the $N-1$th monomer, which we label with subscript $2$ (second particle). (Note that we skip vector notation, as we can always reduce the problem to more scalar degrees of freedom.)\\

\begin{widetext}
\begin{eqnarray}\label{rec1}
 \nonumber
\frac{d^2 R_2^{(N-1)}}{dt^2}(t) &=& -\frac{k}{m}(R_2^{(N-1)}-R_1) -\gamma \frac{dR_2^{(N-1)}}{dt}(t) + \frac{\xi_2(t)}{m} + \frac{\eta^{(N-1)}(t)}{m}  \\
&-&  \int_0^t dt' \mu^{(N-1)}(t-t') \frac{dR_2^{(N-1)}}{dt}(t') + \frac{G^{(N-1)}(t)}{m}
\end{eqnarray}
\end{widetext}
The tagged monomer $R_1$ is attached to $R_2^{(N-1)}$ by a harmonic spring. The force on which is simply given by  
\begin{equation}\label{rec2}
 \Phi^{(N)}(t) = m\frac{d^2 R_1}{dt^2}(t) = -k(R_1 -R_2^{(N-1)}) -m\gamma \frac{dR_1}{dt}(t) + \xi_{1}(t)
\end{equation}
where $\xi_{1}(t)$ and $\xi_{2}(t)$ are the independent white noise on monomers $1$ and $2$.\\ 
These equations represent a system of  2 monomers, one of which is already a coarse grained variable, with memory kernel $\mu^{(N-1)}(t)$, external force $G^{(N-1)}(t)$ and noise $\eta^{(N-1)}(t)$. \\
A second major ingredient in our computation is quite naturally to take the Laplace transform of \eqref{rec1} and \eqref{rec2}.  After integrating out $R_2^{(N-1)}$, we arrive at the following GLE for $R_1$,
 \begin{widetext}
\begin{eqnarray}\label{recgle1}
\nonumber
\tilde{\Phi}^{(N)}(s) &=& -m\kappa\frac{\tilde{\mu}^{(N-1)}(s) +\gamma +s}{ms\tilde{\mu}^{(N-1)}(s) +ms\gamma + ms^2+\kappa}[s\tilde{R}_{1}(s) -R_{1}(0)] - m \gamma[s\tilde{R}_{1}(s) -R_{1}(0)] \\
\nonumber
&+& \frac{ \kappa}{ms\tilde{\mu}^{(N-1)}(s) +ms\gamma + ms^2+\kappa}\tilde{G}^{(N-1)}(s)\\
\nonumber
&+& m\kappa\frac{\tilde{\mu}^{(N-1)}(s) +\gamma +s}{ms\tilde{\mu}^{(N-1)}(s) +ms\gamma + ms^2+\kappa}[R_2^{(N-1)}(0)-R_{1}(0)]\\
\nonumber
&+& \frac{m\kappa}{ms\tilde{\mu}^{(N-1)}(s) +ms\gamma + ms^2+\kappa}\dot{R}_2^{(N-1)}(0)+ \kappa\frac{(\tilde{\eta}^{(N-1)}(s) +\tilde{\xi}_2(s))}{ms\tilde{\mu}^{(N-1)}(s) +ms\gamma + ms^2+\kappa}\\
&+& \tilde{\xi}_{1}(s)
\end{eqnarray}
\end{widetext}
That has to be confronted with \eqref{gle} and in particular with each of the terms on the right-hand side.
In that way we obtain recurrence relations for the memory kernel $\tilde{\mu}^{(N)}(s)$, the noise $\tilde{\eta}^{(N)}(s)$ and the induced force $\tilde{G}^{(N)}(s)$ on the tagged particle when comparing size $N$ polymers with size $N-1$:
 \begin{equation}\label{recmem1}
  \tilde{\mu}^{(N)}(s) = \frac{\kappa(\tilde{\mu}^{(N-1)}(s) +\gamma +s)}{(ms\tilde{\mu}^{(N-1)}(s) +ms\gamma +m s^2+\kappa)}
 \end{equation}
 \begin{equation}\label{recforce1}
 \tilde{G}^{(N)}(s) = \frac{\kappa \tilde{G}^{(N-1)}(s)}{(ms\tilde{\mu}^{(N-1)}(s) +ms\gamma +ms^2+\kappa)}
 \end{equation}
\begin{widetext}
\begin{eqnarray}\label{recnoise1}
 \nonumber
 {\tilde{\eta}}^{(N)}(s) &=& m\tilde{\mu}^{(N)}(s)[R_2^{(N-1)}(0) -R_1(0)] +\frac{m\kappa}{(ms\tilde{\mu}^{(N-1)}(s) +ms\gamma +s^2+\kappa)} \dot{R}_2^{(N-1)}(0) \\
 &+& \frac{\kappa}{(ms\tilde{\mu}^{(N-1)}(s) +ms\gamma +ms^2+\kappa)}(\tilde{\eta}^{(N-1)}(s) +\tilde{\xi}_2(s))
 \end{eqnarray}
\end{widetext}
To these we must add ``initial'' conditions for the recurrence, i.e., to insert the findings for the case $N=2$.  These will enable the correct induction hypothesis.\\ 
Finally, there are the initial conditions to the dynamics; the initial conditions (positions and momenta) of all the other particles (except the tagged particle) contribute to the noise.  Their statistical distribution
is in principle a matter of choice but there are of course dynamically more natural choices.  We will detail them in Appendix \ref{pros}.

\section{Free diffusion under uniform driving}\label{resu}
This and the two following sections summarize the main results of the logic explained in the previous Section.  We always refer to \eqref{gle} for the notation, that we have obtained after integrating out  all but one of the particles.

\subsection{In general}\label{const}
Referring to the dynamics \eqref{us}--\eqref{eq2}--\eqref{cons1}, we define the frequency $\omega$ as $\omega^2 = \frac{\kappa}{m} -\frac{\gamma^2}{4}$.  If $\omega$ is real (inertial case), then the friction 
kernel $\mu^{(N)}$ is oscillating with frequency $\omega$ with decreasing amplitude.  When under high friction, $\omega$ is imaginary (overdamped case), there is monotone decay in time. \\

\vspace{0.2in}

 \begin{figure}[ht!]
 \begin{center}
 \includegraphics[totalheight=6.5cm,width=8cm]{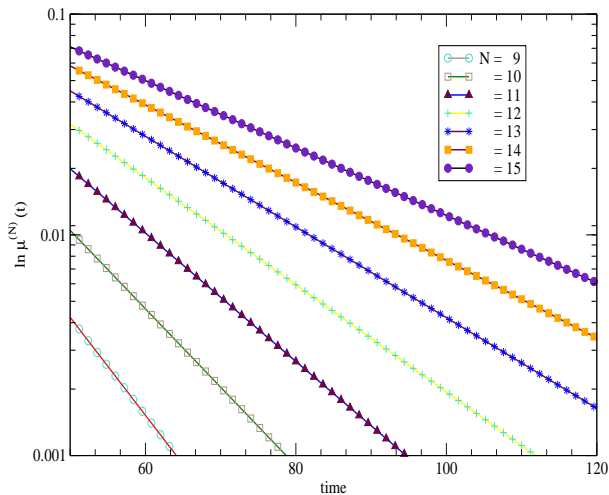}%,width=5in
 \caption{\label{fig3.1}``(Color Online)'' Memory vs time, $\kappa = 3, \xi = 1$.\\
                          The long time limit of memory in a semi-log plot.}
 \end{center}
 \end{figure}

In the long time limit we find that the friction kernel is always exponentially decaying as $\mu^{(N)}(t) \leq e^{-t/\tau_N}$ 
for large times $t$, where the decay time $\tau_N$ is of the order $N^2$ for large $N$, see Fig.~\ref{fig3.1}. In the figures are represented polymers of different sizes. The discrete data points are solutions of numerical evaluation
of analytical results. The lines are fits exhibiting the general nature of these solutions. $\kappa$ is the spring constant and $\chi = m\gamma$. Fig.~\ref{fig3.1} is a semi-log plot, which shows the exponential decay of the friction kernel in long time. The slopes of various lines are proportional to the 
decay exponent $\tau_N$. It is seen in Fig.~\ref{fig3.1} that $\tau_N$ varies as $N^{-2}$, which is already known to hold in full equilibrium \cite{panja2010}.  We show that this remains valid in nonequilibrium.  We have also 
obtained that the time of relaxation is in general bounded from below as $\tau_N > 2/\gamma$, see below in Appendix \ref{mem-rec}. This is indeed clear as the time of relaxation grows with the size
of the polymer and the tagged particle relaxes fastest when it is connected to just one another monomer, $\tau_N$ then being $2/\gamma$.\\

  %\begin{figure}[ht!]
%\begin{center}
%\includegraphics[totalheight=8cm,width=8cm]{shorttimeloglog}%,width=5in
%\caption{\label{fig3.2}The short time behaviour of memory in a log-log plot.  The dotted curves are the analytic values and the full curves are exponential fits.}
%\end{center}
%\end{figure}

As is well known and is also to be expected under nonequilibrium conditions, for short times  $t<\tau_N$, the memory kernel has a power law decay $\mu(t) \propto \frac{1}{t^{1/2}}$, see also \cite{doitheory,de1scaling,panja2010}. Indeed, for short times there is no dependence of memory on system size.\\

Integrating out the other monomers also creates additional coloured noise in the system. The noise $\eta^{(N)}(t)$ is Gaussian with a shifted average and is breaking the second fluctuation--dissipation relation transiently.  
However asymptotically, the stationary covariance
satisfies the second fluctuation-dissipation relation
\begin{equation}\label{FDR}
\lim_{t  \rightarrow \infty} \langle \eta(t+\tau)\eta(t)\rangle = \frac{m}{\beta} \mu(\tau)
\end{equation}
Finally, the external force $f$ on the outer monomer gives rise to a time-dependent statistical force $G(t)$ on the tagged monomer and reaches  exponentially-fast a limiting form. The general behavior is exactly similar to what we make explicit
in the next subsection; see below in Eq. \eqref{force1}.

\subsection{Two monomer case}\label{dimer}
To give immediately more explicit formulae we summarize the results for the dimer-case, which is also used to start the recurrence.  First the memory kernel ($N=2$),
\begin{eqnarray}\label{mem0}
\mu(t) &=& \frac{\kappa}{m} e^{-\gamma t/2}\,[\cos \omega t + \frac{\gamma}{2\omega}\sin \omega t]\\
\nonumber
&\text{ where }& \omega^2 = \frac{\kappa}{m} -\frac{\gamma^2}{4} >0
\end{eqnarray}
For a dimer, a power law decay for small time is not seen, as can be imagined from the fact that a dimer in one dimension has no conformational degrees of freedom. The mean squared end--to--end distance 
$\langle R_{ee}^2\rangle$ is simply equal to the square of the bond length and thus shows only simple diffusion. \\

The coloured noise is
\begin{eqnarray}\label{noise0}
\nonumber
\eta(t) &=& -\kappa[R_1(0)-R_2(0)] [\cos (\omega t) + \frac{\gamma}{2\omega} \sin(\omega t)]e^{-\gamma t/2}\\
\nonumber
&+&  \kappa \dot{R}_2(0)  \frac{\sin(\omega t)}{\omega}e^{-\gamma t/2} \\
&+& \frac{\kappa}{m} \int_0^t  \frac{\sin(\omega(t-t'))}{\omega}  \xi_2(t')e^{-\gamma (t-t')/2} dt'  
\end{eqnarray}
It is natural to take the distribution
\begin{equation}\label{dis1}
\rho_{st}(R_2,\dot{R}_2) = \frac{1}{Z} e^{-\beta H} 
\end{equation}
where $H =  \frac{{m\dot{R}_2}^2}{2} +\frac{\kappa}{2}{(R_1 -R_2)}^2 -f R_2$ that depends on the position $R_1$  of the tagged particle.  When averaged over that initial distribution \eqref{dis1} we get a mean
\[
\langle \eta(t) \rangle = f [\cos (\omega t) +\frac{\gamma}{2\omega} 
\sin( \omega t)] e^{-\gamma t /2}
\]
which is not zero for $f\neq 0$. Indeed, the monomer $R_2$ 
is found more on one side than the other due to the force $f$.\\
The effective force is found to be
\begin{equation}\label{force1}
G(t) = f[1-e^{-\gamma t/2}(\cos(\omega t) +\frac{\gamma}{2\omega} \sin(\omega t))] 
\end{equation} 
exponentially growing to the applied force $f$.  When the force would be time dependent, $f=f_t$, the effective force gets memory and becomes
\[
G(t) = k\int_0^t f_s\,e^{-\gamma(t-s)/2}\,\sin \omega(t-s)\,ds
\] 
That appears to be a general feature of nonequilibrium forcing; they create effective forces that themselves depend on the forcing at all earlier times. We still emphasize  that the nature of nonequilibrium in this paper and in the present example of constant forcing is qualitatively different from the case of driving forces discussed in some previous works \cite{panja2010anom}, polymer translocation
by an external force being one example. Such driving forces are introduced at the macroscopic level or the level of the GLE and hence do not have effect on the nature of the friction kernel or of the noise. On the other hand the nonequilibrium in our work is introduced microscopically, such that
the GLE itself gets modified as a function of the driving.

\subsection{Limiting cases}\label{limit}
\begin{enumerate}
\item  \textbf{The long time limit.}
Motion after $t \gg \tau_N$, of the tagged particle appears to be diffusive with a constant drift $f$.

\item \textbf{Large coupling limit.} 
A large coupling signifies that the restoring force between any two monomers is very strong and hence the monomers undergo high frequency oscillations given by $\omega \simeq \sqrt{\frac{k}{m}} \rightarrow \infty$.
Measurements will typically time--average over a few periods. The time-averaged behavior of the tagged monomer is again of a Brownian particle acted on by a constant force. 
   The effective force goes to the constant force $f$. The time averaged total noise goes to white noise $\xi_1(t)$ 
   and the time averaged memory kernel $\mu(t)$ disappears.  

\item  \textbf{The overdamped limit.}\\
The high friction limit refers to the case when the viscosity of the medium is so high that the acceleration of the monomers is zero, and the only variables which are changing are position.
To take the overdamped limit in a meaningful manner, together with taking the friction coefficient $\gamma$ to infinity one has to take the mass $m$ of all monomers to zero, preserving the product $\chi = m\gamma $  to be
 finite.  The memory kernel then reduces to 
\begin{equation}\label{memdmp}
\mu(t) = ke^{-k t/\chi}
\end{equation} 
The memory kernel after taking the continuum limit is 
\begin{equation}\label{memdeb}
\mu(t) = 2\sqrt{\frac{\pi \chi k}{t}}e^{-t/\tau}
\end{equation}
where $\tau = N^2 \chi/(\pi^2 k)$, as shown before \cite{panja2010} under equilibrium dynamics. 
\end{enumerate} 

\section{Non-uniform driving}\label{she-dimer}
We are now in two dimensions with forcing at one end of the polymer in the horizontal direction with an amplitude that is proportional to the vertical distance.  That dependence is similar to a shearing force, but we do not insist here on the presence of a fluid (as we treat the Rouse model and not e.g. the Zimm model).  Rather, we have in mind that we can manipulate  the outer monomer of a polymer in a melt in a non-uniform way.  To a good approximation, that would be the case when nanoparticles are attached to the polymer and undergo non-rigid rotation, where the angular velocity depends on the radial distance (here, the vertical distance).\\ 
An interesting result here is that the friction kernel $\mu^{(N)}(t)$  is identical to the case of constant forcing (previous section).  To be explicit and without loss of essential information we can already state the results for $N=2$.\\
 The memory kernel in each direction is given by
\[ \mu_{1x}(t)= \mu_{1y}(t) = \frac{\kappa}{m} e^{-\gamma t/2}[\cos(\omega t) + \frac{\gamma}{2\omega} \sin(\omega t)] \] which is indeed the same as for a dimer in free space under constant forcing. That is due to the fact that the external forcing does not couple to velocity but only to position. 
The nature of $\mu$ in the two mutually perpendicular directions and various limits  hence remains the same. \\

The nature of the induced noise however gets modified due to the different nature of the external force. The noise in general is dependent on the initial positions and velocities of all the monomers, and hence picks up additional 
contributions from the external force which is coupled to the $y$ component of the position of the first monomer. Here is the explicit noise function\\

\begin{widetext}
\begin{eqnarray}\label{noise1x}
\nonumber
\eta_{1x}(t) &=& -\kappa[R_{1x}(0)-R_{2x}(0)] (\cos (\omega t) + \frac{\gamma}{2\omega} \sin(\omega t))e^{-\gamma t/2}\\
\nonumber
&+&  \kappa \dot{R}_{2x}(0) e^{-\gamma t/2} \frac{\sin(\omega t)}{\omega} \\
\nonumber
&+& \frac{\kappa}{m} \int_0^t  e^{-\gamma (t-t')/2} \frac{\sin(\omega(t-t'))}{\omega}  \xi_{2x}(t') dt' \\
\nonumber
&+& \frac{f}{m \omega} \kappa\int_0^t dt' e^{-\gamma(t-t')/2} sin(\omega (t-t'))[R_{2y}(0)(\cos(\omega t')+ \\
\nonumber
&& \frac{\gamma}{2\omega}\sin(\omega t')) + \dot{R}_{2y}(0) \frac{\sin(\omega t')}{\omega}] \\
&+& \frac{f}{m\omega}\frac{\kappa}{m \omega} \int_0^t \int_0^{t'} dt' dt'' e^{-\gamma(t-t'')/2} \sin(\omega (t-t'))\sin(\omega (t'-t'')) \xi_{2y}(t'')
\end{eqnarray}
\end{widetext}
Putting $f =0$ in \eqref{noise1x} gives us back the noise on a polymer under constant force \eqref{noise0}. The external force couples the $x$ component of noise to the dynamics in $y$ direction. 
The initial positions in the $y$-direction as well as the component of the white noise $\xi_{2y}(t)$ in the $y$-direction now play a role in the dynamics in the $x$-direction of the second monomer. 
The $y$-component of the noise remains unaffected by the force, since the external force does not couple to the motion in the $y$-direction.
\begin{eqnarray}\label{noise1y}
\nonumber
\eta_{1y}(t) &=& -\kappa[R_{1y}(0)-R_{2y}(0)] (\cos \omega t + \frac{\gamma}{2\omega} \sin \omega t)e^{-\gamma t/2}\\
\nonumber
&+&  \kappa \dot{R}_{2y}(0) e^{-\gamma t/2} \frac{\sin(\omega t)}{\omega} \\
\nonumber
&+& \frac{\kappa}{m} \int_0^t  e^{-\gamma (t-t')/2} 
\frac{\sin(\omega(t-t'))}{\omega}  \xi_{2y}(t') dt' \\
\end{eqnarray}
Now we come to the induced force. The $x-$component of the effective force is
\begin{eqnarray}\label{force1x}
\nonumber
G_{1x}(t) &=& f\frac{\kappa^2}{m^2\omega^2} \int_0^t dt' \int_0^{t'} dt'' e^{-\gamma(t-t'')/2} \sin(\omega(t-t'))\\
&&\sin(\omega(t'-t'')) R_{1y}(t'')
\end{eqnarray}
Again, as a nonequilibrium effect, the effective force has memory.  On the other hand its $y$-component $G_{1y}(t) $ stays zero. Indeed, since the applied force itself is acting in the $x$-direction, there is no reason why the 
effective dynamics in the $y$-direction of the tagged monomer should get affected by it.

Let us now go to the results for general $N$. The memory kernels $\tilde{\mu}_{x}^{(N)}(s)$ and $\tilde{\mu}_{y}^{(N)}(s)$ look as they did in the case of constant force \eqref{recmem1}, with the same ``initial'' conditions.\\
The recurrence relation for the coloured noise picks up changes due to shearing, similar as discussed under \eqref{noise1x} and \eqref{noise1y}; they arise due to the fact that the non-uniform forcing couples the $x$ and $y$ components of motion.\\

\begin{widetext}
\begin{eqnarray}\label{rec-noise1x}
\nonumber
\tilde{\eta}_x^{(N)}(s) &=& m\tilde{\mu}_x^{(N)}(s)[R_x^{(N-1)}(0) -R_x(0)] +\frac{m\kappa}{(ms\tilde{\mu}_x^{(N-1)}(s) +ms\gamma +s^2+\kappa)} \dot{R}_x^{(N-1)}(0) \\
\nonumber
 &+& \frac{\kappa}{(ms\tilde{\mu}_x^{(N-1)}(s) +ms\gamma +ms^2+\kappa)}(\tilde{\eta}_x^{(N-1)}(s) +\tilde{\xi}_x^{(N-1)}(s))\\
&+& f \kappa^3\frac{\{[s+\gamma +\tilde{\mu}_y^{(N-1)}(s)]R_y^{(N-1)}(0) + \dot{R}_y^{(N-1)}(0) + \tilde{\xi}_y^{(N-1)}(s)/m + \tilde{\eta}_y^{(N-1)}(s)/m\}}{[ms^2 + m\gamma s +\kappa +ms\tilde{\mu}_x^{(N-1)}][ms^2 + m\gamma s +\kappa +ms\tilde{\mu}_y^{(N-1)}][ms^2 + m\gamma s +\kappa]}
\end{eqnarray}
\end{widetext}
The most interesting aspect of the non-uniform case is the appearance of memory in the induced force. We have already seen this in the case of two-monomers \eqref{force1x}. This behaviour persists in general with,
\begin{equation}\label{rec-she-force-x}
 \tilde{F}_x^{(N)}(s) = \tilde{G}_x^{(N)}(s) \tilde{R}_y^{(N)}(s)
\end{equation}
\begin{widetext}
\begin{equation}\label{ker}
       \tilde{G}_x^{(N)}(s) = \frac{\kappa^2 }{ [ms^2 +\gamma s +k +ms\tilde{\mu}_x^{(N-1)}(s)][ms^2 +\gamma s +k +ms\tilde{\mu}_y^{(N-1)}(s)]} \tilde{G}_x^{(N-1)}(s)
      \end{equation}
\end{widetext}
Starting the recurrence with  a single monomer where $\tilde{G}_x^{(1)}(s) = f$, all subsequent forces can be determined using equations \eqref{rec-she-force-x}-- \eqref{ker}. The initial force on a single monomer in the $y$-direction is zero, hence  $\tilde{F}_y^{(N)}(s) = 0$ as also seen in the two monomer case.\\
We studied the asymptotic behaviour of the force-memory kernel $\tilde{G}_x^{(N)}(s)$ in the same spirit as in Appendix \ref{mem-rec}. It can be shown easily following the same line of arguments that $\tilde{G}_x^{(N)}(s)$ decays exponentially in time.  
In the long time limit, for all $N$,  
\[ G_x^{(N)}(t) < e^{-\gamma t/2} \]

\section{Trapped monomer}\label{trap}

Upon introducing an external potential $U_{\text{ext}}$ such that the force in the $x$-direction on the outer edge depends on the $x$-component of the distance of the monomer from a fixed origin, the resulting force is not non-conservative but simply trapping. That is thus an equilibrium reference; the force is conservative in nature. The result a rescaling of the frequency $\Omega^2 = \frac{k+f}{m}-\frac{\gamma^2}{4}$. 
 The effective force on the tagged monomer due to the action of this external potential is
\begin{equation}
                       F(t)= -\frac{\kappa f}{\kappa+f} R_1(t) + \frac{\kappa fQ}{\kappa +f}\{1-e^{-\gamma t/2}(\cos \Omega t +\frac{\gamma}{2\Omega}\sin{\Omega t})\}
                      \end{equation}
We recognize the effective spring replacing two springs connected in series,
\[\frac{1}{\kappa_{eff}} = \frac{1}{\kappa} +\frac{1}{f}\]
This is another way to understand the net restoring force on $R_1$. After all it looks like a trapping potential around the origin but of strength $\kappa_{eff}$. \\
The coloured noise due to unknown initial conditions is
\begin{eqnarray}
\nonumber
\eta_1(t) &=&  \frac{\kappa^2}{\kappa+f}(R_2(0) - R_1(0))e^{-\gamma t/2}(\cos \Omega t +\frac{\gamma}{2\Omega}\sin{\Omega t})\\
\nonumber
& +& \frac{\kappa f}{k+f}R_2(0)e^{-\gamma t/2}(\cos \Omega t +\frac{\gamma}{2\Omega}\sin{\Omega t}) \\
\nonumber
&+& \kappa \dot{R}_2(0)e^{-\gamma t/2} \frac{\sin \Omega t}{\Omega} \\
&+& \frac{\kappa}{m} \int_0^t e^{\gamma(t-t')/2}\xi_2(t') \frac{\sin(\Omega(t-t'))}{\Omega} dt'
\end{eqnarray}
where $\langle \eta_1(t) \rangle_{\rho_{st}}^{R_1} = \frac{\kappa fQ}{\kappa+f} e^{-\gamma t/2}(\cos(\Omega t) +\frac{\gamma}{2\Omega}\sin{\Omega t}) $ 
for distribution
\[
\rho_{st}(R_2) =\frac{1}{Z} e^{-\beta\frac{(\kappa+f)}{2}(R_2-\frac{(\kappa R_1 +fQ)}{\kappa+f})^2} e^{\beta \frac{(\kappa R_1 +fQ)^2}{2(\kappa+f)}} e^{-\beta\frac{(\kappa R_1^2 +fQ^2)}{2}}
\]
The memory kernel is given as
\begin{equation}
 \mu(t) = \frac{k^2}{m(k+f)}e^{-\gamma t/2}(\cos{\Omega t} +\frac{\gamma}{2\Omega}\sin{\Omega t})
 \end{equation}

 Given the conservative nature of the forces, the second fluctuation-dissipation theorem is seen to hold:

\[\langle\eta_1(t_1)\eta_1(t_2)\rangle_{\rho} = \frac{m}{\beta}\mu(\tau)\]

where $\tau = t_1 -t_2$ and $t_1+t_2 \longrightarrow \infty$.

\section{Conclusions and outlook}
Integrating out degrees of freedom introduces
non-Markovian noise, effective forces and memory in a tagged particle dynamics. That is true in equilibrium as in nonequilibrium, and, when starting from coupled diffusion processes, the result is a generalized Langevin equation.  Certain more detailed aspects are also unchanged, like the anomalous nature of the memory kernel for short times which goes into pure diffusion for long times, or the $N^2$ dependence of the relaxation times.  Other important aspects fundamentally change when the integration is over nonequilibrium degrees of freedom.  Naturally, the remaining and visible degrees of freedom inherit nonequilibrium features and detailed balance gets broken.  As a result, the so called second fluctuation-dissipation theorem or Einstein relation gets violated.  For the moment however, there is no systematic understanding of exactly {\it how} that Einstein relation is modified.  To put it differently, when considering a diffusion model for a particle (e.g. colloids) in a nonequilibrium environment such as the visco-elastic medium of the cell, we have little idea of how to relate the noise with the friction term, be that they have the same physical origin \cite{bohec2012probing}.   The outlook is then to find the analogue of what has been called the frenetic contribution to the first fluctuation-dissipation theorem \cite{baiesi2012update}.
Indeed we expect a non-entropic and more kinetic contribution in the breaking of the second fluctuation-dissipation theorem, much like discussed for the modification of the Sutherland-Einstein relation \cite{maes2012fluctuation}.  For the moment however, we must deal with examples and prototypical examples, such as the Rouse model of the present paper, where exact computations are possible.  There indeed, say in the case of non-uniform driving, the second fluctuation-dissipation relation is broken, but for the uniform driving that is only a transient effect as found in \eqref{FDR}. A more general theory will of course need to conform to the findings of the present paper.\\
A second set of more general research questions really inverts the calculations of the present paper.  The aim is then to be able to reconstruct the nonequilibrium forcing on the hidden degrees of freedom from the effective motion of the probe or tagged or visible degrees of freedom.
The standard example from equilibrium statistical mechanics is the free energy of a thermodynamic system which can be measured from the work on some probe that is coupled to the system.  For nonequilibrium statistical mechanical systems there are plenty of nonequilibrium entropies and fluctuation functionals, \cite{maes2012nonequilibrium} but so far, no solid and general operational meaning has been attached to them. We would again like to determine these nonequilibrium fluctuation functionals from the effective forces on probes.  In the present paper it would mean to reconstruct important nonequilibrium features of the full polymer dynamics from the motion and effective dynamics of the tagged monomer. Clearly, before that program can start, the direct question as in the paper must be sufficiently understood.  We conclude that the Rouse dynamics provides an interesting and important playground for questions that in the future must be addressed in the construction of a nonequilibrium statistical mechanics.\\

\noindent {\bf Acknowledgment:}  We are grateful to D. Panja, C. Vanderzande, A. Salazar and M. Baiesi for interesting suggestions.\\

\appendix
\section{Details and computations}\label{pros}
 Let us specify to the case of constant forcing.  The initial conditions for the recurrence are
$\tilde{\mu}^{(1)}(s) =0; \text{  } \tilde{G}^{(1)}(s) = f/s \text{ ; } \tilde{\eta}^{(1)}(s) =0 $.\\
The stationary distribution of $R_2^{(N-1)}(0)$ given $R_{1}(0)$ is given by
\[\rho_{st} = \frac{1}{Z} e^{-\beta H_{st} }\]
where \[H_{st} = \frac{\kappa}{2}[R_2^{(N-1)}(0)-R_1(0)]^2 - G_{st}^{(N-1)}R_2^{(N-1)}(0) \]
where we take it that external force has always been on.  Of course, it is also important here to separate transient from stationary behavior.  For example, for the force on $q_2$, we have at stationarity $G_{st}^{(N-1)} = \lim_{t\rightarrow \infty} G^{(N-1)}(t)$ and as an illustration we will show using recurrence that in the case of constant forcing it always equals the originally applied force $G_{st}^{(N-1)} = f$.\\

We start from the relation
\[
\lim_{t\longrightarrow \infty} G^{(N-1)}(t) = \lim_{s\longrightarrow 0} s\tilde{G}^{(N-1)}(s)\]
The recurrence relation for the force \eqref{recforce1} starts from a dimer,
\begin{eqnarray}
 \nonumber
\tilde{G}^{(2)}(s) &=&  \frac{\kappa \tilde{G}^{(1)}(s)}{(ms\tilde{\mu}^{(1)}(s) +ms\gamma +ms^2+\kappa)}\\
\nonumber
&=& \frac{\kappa f}{s(ms\gamma +ms^2+\kappa)}
\end{eqnarray}
By a simple calculation it is seen that \[\lim_{s\longrightarrow 0} s\tilde{G}^{(2)}(s) = f\]
If now for a polymer of size $N-2$ (induction hypothesis)  
\[
\lim_{s\longrightarrow 0} s\tilde{G}^{(N-2)}(s) = f\]
We can use the recurrence relation and the property $\lim_{s\longrightarrow 0} s\tilde{\mu}^{(N)}(s)=0$ shown in \ref{mem-rec}, to see that also
\begin{eqnarray}\label{forceasymp}
 \nonumber
\lim_{s\longrightarrow 0} s\tilde{G}^{(N-1)}(s) = f
\end{eqnarray}
as wanted.   The stationary distribution thus is given by
\begin{equation}\label{recrho}
 \rho_{st} = \frac{1}{Z} e^{-\beta(\frac{\kappa}{2}(R_2^{(N-1)}(0)-R_{1}(0)^2) -f R_2^{(N-1)}(0)) }
\end{equation}
The mean noise is
 \[\langle \tilde{\eta}_{1}^{(N)}(s) \rangle_{st} = m\frac{f}{\kappa}\tilde{\mu}_{1}^{(N)}(s)\]

\subsubsection{Asymptotic behaviour of memory}\label{mem-rec}
We show here that the memory kernel $\tilde{\mu}^{(N)}(s)$ decays exponentially in the long time limit, again by recurrence.  We take the constant force case as simplest example.\\
From \eqref{mem0} we see that for a dimer
 \[
 \lim_{t\rightarrow \infty} e^{\lambda t} \mu^{(2)}(t) =  0 \text{ for $\lambda < \gamma/2$ }\] which translates to 
\begin{equation}\label{dec1}
\lim_{s\rightarrow 0} s \tilde{\mu}^{(2)}(s-\lambda) = 0
\end{equation} in the Laplace space.\\
Let us assume that for a polymer of size $N-1$ 
\begin{equation}\label{dec2}
\lim_{s\rightarrow 0} s \tilde{\mu}^{(N-1)}(s-\lambda) = 0 \,\text{  (induction hypothesis) }
\end{equation}
and let us choose $\lambda =\gamma/4$.
That would show that for a polymer of size $N$,\\

 \begin{widetext}
 \begin{eqnarray}\label{memr1}
 \nonumber
 \lim_{s\rightarrow 0} s\tilde{\mu}^{(N)}(s-\gamma/4) &=& \lim_{s\rightarrow 0} s\frac{\kappa(\tilde{\mu}^{(N-1)}(s-\gamma/4) + \gamma + s -\gamma/4)}{(m(s-\gamma/4) \tilde{\mu}^{(N-1)}(s-\gamma/4) + m(s-\gamma/4)\gamma +m(s-\gamma/4)^2 +\kappa)}\\
 &=& \lim_{s\rightarrow 0} 3s\gamma/4 = 0
 \end{eqnarray}
\end{widetext}
where we have used recurrence relation \eqref{recmem1} and hypothesis \eqref{dec2} and the fact that $\lim_{s\rightarrow 0} \tilde{\mu}^{(N)}(s)$ is a constant, as is easy to show.  The above result translates to \[
\lim_{t\rightarrow \infty} e^{\gamma t/4 } \mu^{(N)}(t) =  0\]
 Hence  in the long time limit, for all $N$,  \[
 \mu^{(N)}(t) < e^{-\gamma t/2} \] which also proves the claim made in \ref{const} that the time of relaxation is bounded from below by $2/\gamma$.  
 
\subsubsection{Second fluctuation-dissipation relation in Laplace Space}\label{FDR-rec}
The second fluctuation--dissipation relation says that the stationary noise auto-correlation function is proportional to the memory kernel $\mu(t)$ through inverse temperature $\beta$,
\begin{equation}\label{FDR2}
\langle\eta(t)\eta(t+\tau)\rangle = \frac{m}{\beta} \mu(\tau) + O(\frac{1}{t},\tau)
\end{equation}
 such that in the long time limit all terms of the order $1/t$ or greater drop out. \\
We continue by deriving that relation in Laplace space. Let $s$ be the variable in the Laplace space, domain $|Re\{s\}| < \gamma/2$, such that the Laplace transform is well defined; see Appendix  \ref{mem-rec}. \\

\begin{widetext}
 \begin{eqnarray}\label{lap}
 \nonumber
 \langle \tilde{\eta}(s)\tilde{\eta}(s')\rangle &=& \int_0 ^\infty  \int_0 ^\infty  e^{-st} e^{-s't'}\langle \eta(t) \eta(t')\rangle dt dt' \\
 \nonumber
 \text{Let } t' = t+\tau \\
  \nonumber
 &=& \int_0^\infty  e^{-st} dt \int_{-t}^\infty  e^{-s't}e^{-s'\tau}\langle \eta(t)\eta(t+\tau)\rangle d\tau\\
 \nonumber
 &=& \int_0^\infty  e^{-(s+s')t} dt \int_{-t}^\infty  e^{-s'\tau}\langle \eta(t)\eta(t+\tau)\rangle d\tau\\
 \text{Let } (s+s')t = T\\
\nonumber
  &=& \frac{1}{s+s'} \int_0^\infty  e^{-T} dT \int_{-\frac{T}{s+s'}}^\infty e^{-s'\tau}\langle \eta(\frac{T}{s+s'})\eta(\frac{T}{s+s'} +\tau)\rangle d\tau
 \end{eqnarray}
 \end{widetext}
Rewriting \eqref{FDR2} and plugging in the result in \eqref{lap}
  \begin{equation}
   \nonumber
\langle\eta(\frac{T}{s+s'})\eta(\frac{T}{s+s'}+\tau)\rangle = \frac{m}{\beta} \mu(\tau) + O(s+s')
  \end{equation}
\begin{widetext}
 \[\langle \tilde{\eta}(s)\tilde{\eta}(s')\rangle = \frac{1}{s+s'} \int_0^\infty e^{-T} dT \int_{-\frac{T}{s+s'}}^\infty e^{-s'\tau}(\frac{m}{\beta} \mu(\tau) + O(s+s')) d\tau\]
\end{widetext}
  Hence,
  \begin{eqnarray}\label{FDR3}
 \nonumber
 \lim_{s+s' \rightarrow 0} (s+s') \langle\tilde{\eta}(s)\tilde{\eta}(s')\rangle &=& \frac{m}{\beta}\int_{-\infty}^\infty e^{-s'\tau}  \mu(\tau) d\tau\\
 &=&  \frac{m}{\beta}(\tilde{\mu}(s) +\tilde{\mu}(-s))
 \end{eqnarray}
is the form of the second fluctuation-dissipation relation in Laplace space.

\subsubsection[]{Proof by induction}
We prove our claim that the second fluctuation-dissipation relation holds in case of constant force, for a polymer of general size $N$.  We give the explicit calculation for the case of a dimer.\\
Given memory kernel \eqref{recmem1}  and noise \eqref{recnoise1}, which is distributed as $\rho_{st}(R_2,\dot{R}_2)$ as given in \eqref{recrho},
%\[\tilde{\mu}_2(s) = \frac{k(s+\gamma)}{m^2s(s+\gamma/2)^2+\omega^2}\]
%\[\tilde{\mu}_2(s) +\tilde{\mu}_2(-s) = \frac{2 \gamma k^2}{m^2\{(s-\gamma/2)^2+\omega^2\}\{(s+\gamma/2)^2+\omega^2 \}}\]
one can calculate the correlation function $\langle \tilde{\eta}^{(2)}(s) \tilde{\eta}^{(2)}(s')\rangle_{st}$.  Using the relation 
\begin{equation}\label{corwhite}
 \lim_{s+s' \rightarrow 0} (s+s')\langle \tilde{\xi}_i(s)\tilde{\xi}_j(s')\rangle = \frac{2m\gamma}{\beta}\delta_{ij}
\end{equation}
it is shown by a simple calculation that 

\begin{equation}
%lim_{s+s' \rightarrow 0} (s+s') \langle\tilde{\eta}(s)\tilde{\eta}(s')\rangle &=& \frac{k^2 2m\gamma}{\beta m^2((s+\gamma/2)^2 +\omega^ 2)((s-\gamma/2)^2 +\omega^ 2)} \\
\lim_{s+s' \rightarrow 0} (s+s') \langle\tilde{\eta}^{(2)}(s)\tilde{\eta}^{(2)}(s')\rangle = \frac{m}{\beta}\{\tilde{\mu}_2(s) +\tilde{\mu}_2(-s)\}
\end{equation}
which proves the result.\\ To prove it for a general polymer, we use the induction hypothesis  that for a polymer of size $N-1$ the second fluctuation-dissipation relation holds:\\

 \begin{widetext}
\[
\lim_{s+s' \rightarrow 0}(s+s') \langle \tilde{\eta}^{(N-1)}(s)\tilde{\eta}^{(N-1)}(s')\rangle_{st} = \frac{m}{\beta}\{\tilde{\mu}^{(N-1)}(s) +\tilde{\mu}^{(N-1)}(-s)\}\]
\end{widetext}
From the recurrence relations \eqref{recnoise1} and from \eqref{corwhite}, one easily shows that

\begin{widetext}
\begin{eqnarray}
\nonumber
\lim_{s+s' \rightarrow 0}(s+s') \langle \tilde{\eta}^{(N)}(s)\tilde{\eta}^{(N)}(s')\rangle_{st} 
&=& \frac{\kappa^2(\tilde{\mu}^{(N-1)}(s) +\tilde{\mu}^{(N-1)}(-s))}{m\beta (s\tilde{\mu}^{(N-1)}(s) +s\gamma +s^2 +\frac{\kappa}{m})(-s\tilde{\mu}^{(N-1)}(-s) -s\gamma +s^2 +\frac{\kappa}{m})} \\
\nonumber
&+& \frac{2\gamma \kappa^2}{m\beta (s\tilde{\mu}^{(N-1)}(s) +s\gamma +s^2 +\frac{\kappa}{m})(-s\tilde{\mu}^{(N-1)}(-s) -s\gamma +s^2 +\frac{\kappa}{m})}
\end{eqnarray}
\end{widetext}

Using the recurrence relations for memory \eqref{recmem1},\\

\begin{widetext}
\begin{eqnarray}
\nonumber
\tilde{\mu}^{(N)}(s) +\tilde{\mu}^{(N)}(-s) &=& \frac{\kappa^2(\tilde{\mu}^{(N-1)}(s) +\tilde{\mu}^{(N-1)}(-s))}{m^2(s\tilde{\mu}^{(N-1)}(s) +s\gamma +s^2+\frac{\kappa}{m})(-s\tilde{\mu}^{(N-1)}(-s) -s\gamma +s^2+\frac{\kappa}{m})}\\
\nonumber
&+& \frac{2\kappa^2\gamma}{m^2(s\tilde{\mu}^{(N)}(s) +s\gamma +s^2+\frac{\kappa}{m})(-s\tilde{\mu}^{(N)}(-s) -s\gamma +s^2+\frac{\kappa}{m})}
\end{eqnarray}
\end{widetext}
which proves the claim.
Therefore, the second fluctuation-dissipation relation holds for a polymer of arbitrary size under the action of a constant force.\\

In the case of non-uniform forcing, we are in two dimensions and the computations become more involved, but the basic recurrence relations remain in place.

\bibliography{entropicforces}

\end{document}